\documentclass[journal]{IEEEtran}
\IEEEoverridecommandlockouts
\usepackage{cite}
\usepackage{booktabs}
\usepackage{amsmath,amssymb,amsfonts}
\ifCLASSINFOpdf
\usepackage{graphicx}
\usepackage[dvipsnames,table,xcdraw]{xcolor}
\usepackage{siunitx}
\usepackage{amsfonts}
\usepackage{amssymb,mathtools}
\usepackage{algpseudocode}
\usepackage[linesnumbered,ruled,vlined]{algorithm2e}
\usepackage{nomencl}
\usepackage{graphicx}
\usepackage{textcomp}
\usepackage{float}
\usepackage{siunitx}
\usepackage{tabularx}
\usepackage{algpseudocode}
\usepackage{capt-of}  
\usepackage{cuted}
\usepackage{svg}
\usepackage{enumitem}
\usepackage{mdframed}
\usepackage{tcolorbox}
\usepackage{multirow}
\usepackage[acronym, nonumberlist, nopostdot, nogroupskip]{glossaries-extra}

\usepackage{caption}

\usepackage{subcaption}
\usepackage{comment}
\newcommand{\RNum}[1]{\uppercase\expandafter{\romannumeral #1\relax}}

\def\BibTeX{{\rm B\kern-.05em{\sc i\kern-.025em b}\kern-.08em
    T\kern-.1667em\lower.7ex\hbox{E}\kern-.125emX}}
    
\usepackage{atbegshi}
\AtBeginDocument{\AtBeginShipoutNext{\AtBeginShipoutDiscard}}

\makenoidxglossaries
\setabbreviationstyle[acronym]{long-short}

\newacronym{ai}{AI}{Availability Index}
\newacronym{ac}{AC}{Alternating Current}
\newacronym{bess}{BESS}{Battery Energy Storage System}
\newacronym{cfd}{CFD}{Computational Fluid Dynamics}
\newacronym{doc}{DOC}{Depth of Cycle}
\newacronym{ecm}{ECM}{Equivalent Circuit Model}
\newacronym{ems}{EMS}{Energy Management System}
\newacronym{fdm}{FDM}{Finite Difference Method}
\newacronym{ff}{FF}{Fulfilment Factor}
\newacronym{fcr}{FCR}{Frequency Containment Reserve}
\newacronym{ftcs}{FTCS}{Forward Time Central Space}
\newacronym{kpi}{KPI}{Key Performance Indicator}
\newacronym{lfp}{LFP}{Lithium Iron Phosphate}
\newacronym{lp}{LP}{Linear Programming}
\newacronym{li-nmc}{Li-NMC}{Lithium Nickel Manganese Cobalt Oxide}
\newacronym{lut}{LUT}{Look Up Table}
\newacronym{miqp}{MIQP}{Mixed Integer Quadratic Programming}
\newacronym{minlp}{MINLP}{Mixed Integer Non Linear Programming}
\newacronym{moo}{MOO}{Multi Objective Optimization}
\newacronym{mpc}{MPC}{Model Predictive Control}
\newacronym{nlp}{NLP}{Nonlinear Programming}
\newacronym{ocv}{OCV}{Open Circuit Voltage}
\newacronym{pi}{PI}{Proportional Integral}
\newacronym{sci}{SCI}{Self Consumption Increase}
\newacronym{soc}{SOC}{State of Charge}
\newacronym{soh}{SOH}{State of Health}

\begin{document}





\title{Multi-Objective Nonlinear Power Split Control For BESS With Real-Time Simulation Feedback}

\thanks{The work of Vivek Teja Tanjavooru was funded by Bayerische Forschungstiftung under the research project KIM-Bat.}

\author{
  \IEEEauthorblockN{
    Vivek Teja Tanjavooru\textsuperscript{1,2},
    Prashant Pant\textsuperscript{2},
    Thomas Hamacher\textsuperscript{2},
    Holger Hesse\textsuperscript{1}
  }

  \IEEEauthorblockA{\textsuperscript{1}\textit{Institute of Energy and Drive Technology, University of Applied Sciences Kempten}, Germany}
  \\
  \IEEEauthorblockA{\textsuperscript{2}\textit{Center for Combined Smart Energy Systems (CoSES), Technical University of Munich}, Germany}
  \\
  \IEEEauthorblockA{
    \{vivek.tanjavooru, holger.hesse\}@hs-kempten.de,
    \{prashant.pant, thomas.hamacher\}@tum.de
  }
}

\maketitle

\setcounter{page}{1}
\begin{abstract}

This paper presents a mixed-integer, nonlinear, multi-objective optimization strategy for optimal power allocation among parallel strings in Battery Energy Storage Systems (BESS). High-fidelity control is achieved by co-simulating the optimizer with a BESS electro-thermal simulation that models spatial thermal dynamics of the battery, providing real-time State of Charge (SOC) and temperature feedback. The optimizer prioritizes reliability by enforcing power availability as a hard constraint and penalizing battery thermal derating. Within these bounds, the controller performs a Pareto sweep on the relative weights of inverter and battery losses to balance the trade-off between inverter efficiency and battery efficiency. The inverter loss model is based on an empirical lookup table (LUT) derived from a commercial inverter system, while the battery thermal loss model uses SOC and temperature-dependent internal resistance, with electric current computed from the battery Equivalent Circuit Model (ECM). When the optimization was applied to a two-string BESS, the competing effects of inverter and battery losses on system availability and thermal derating were observed. The balanced operation yielded improvements of 1\% in battery efficiency, 1.5\% in inverter efficiency, and 2\% in derating efficiency, while maintaining higher availability. Additionally, a \SI{5}{\degreeCelsius} reduction in BESS peak temperature also suggests reduced thermal stress without compromising availability.
\end{abstract}

\begin{IEEEkeywords}
Battery Energy Storage Systems (BESS), Multi-Objective Optimization, Efficiency, Thermal Simulation, Inverter Loss Model, Power Derating, Power Split Control.
\end{IEEEkeywords}
\vspace{-2mm}
\printnoidxglossary[type=acronym, title={List of Abbreviations}]

\section {Introduction}
The global shift toward decentralized renewable energy, alongside the liberalization of electricity markets, has positioned \gls{bess} as a pivotal technology. Their capability to deliver energy at variable ramp rates enables effective participation across multiple market segments, including energy arbitrage and \gls{fcr} trading \cite{Pant_ancillary}. To fulfill these roles, \gls{bess} must maintain high-power operation over extended periods, which elevates internal temperatures and can trigger thermal derating \cite{RUAN2023232805}. This derating not only limits the system’s ability to meet grid support obligations but also diminishes potential revenue \cite{derate,Schimpe_2021}. Moreover, repeated thermal stress accelerates battery degradation, reducing capacity retention, power output, and overall system reliability over time \cite{Tanjavooru.2025}.

To address these challenges, control strategies must be system-aware. Advances in battery management systems, especially for electric vehicles, highlight the effectiveness of optimal control methods such as \gls{mpc} in regulating charge/discharge, temperature, \gls{soc}, and \gls{soh}. Variants including nonlinear, linear, explicit, and hierarchical \gls{mpc} have demonstrated strong potential to improve safety and efficiency while managing computational complexity \cite{Kim.2024,Zou.2018}. Additionally, \gls{moo}, commonly used in electric vehicles to balance fuel consumption, battery usage, and emissions offers a useful framework that can be adapted to \gls{bess} for optimizing power flow, thermal load, and degradation \cite{Du.2021}.

Extending these optimal control approaches to \gls{bess} with multiple strings/ packs and integrated inverters in grid applications increases complexity, as power distribution must account for non-uniform capacities, thermal behavior, and degradation dynamics under demanding operational constraints \cite{Li.2016}. Controllers must ensure high system availability while simultaneously managing non-uniform temperature evolutions across battery modules, which contribute to uneven aging and potential premature degradation. At the same time, maximizing overall system efficiency remains a critical goal. However, these objectives often conflict with one another, requiring techniques such as Pareto analysis to assess the trade-offs \cite{Jin.2024}. 

For instance, operating the battery aggressively at nominal power levels to maximize inverter efficiency can elevate internal temperatures, thereby accelerating battery aging. While these higher temperatures may initially improve battery efficiency by reducing internal resistance, sustained thermal stress ultimately degrades battery health \cite{Ma.2018}. Conversely, strict thermal management aimed at limiting aging often reduces available power and system up-time. Consequently, effective control strategies must carefully balance these competing priorities to achieve an optimal trade-off among reliability, longevity, and overall system performance.

Accurate estimation of battery cell states and their temporal evolution inherently requires nonlinear modeling. Prior studies integrating such nonlinearities within optimization frameworks have demonstrated clear advantages in enhancing controller awareness of the actual system state compared to simplified linear approximations \cite{yin2025evaluating}. Notably, \gls{miqp} has been applied to establish aging-aware energy management \cite{Kumtepeli.2019b}, and benchmarking studies comparing \gls{nlp} with \gls{lp} have shown improved capture of complex battery dynamics \cite{cornejo2025evaluating}. 

Extending prior studies, this work introduces a \gls{minlp}-based \gls{mpc} framework tailored to \glspl{kpi} relevant to multi-string \gls{bess} operation. The proposed formulation enables the \gls{moo} of minimizing a quadruple set of objectives: (i) system availability, (ii) temperature-induced derating losses, (iii) inverter efficiency, and (iv) battery efficiency. Given the critical role of grid support and revenue generation in grid-connected applications, system availability is prioritized as the primary objective within this control framework.

\subsection{Identified research gap}
A systematic review of literature on power split control for parallel BESS strings, synthesized in Table \ref{tab:synthesis_bullets}, reveals several critical research gaps despite the growing body of work:

\begin{enumerate}
    \item The use of multi-objective \gls{mpc} control for power allocation in a multi-string \gls{bess} is limited, with most works focusing only on single objective formulations.
    \item Integrated modeling of thermal derating, inverter and battery efficiencies, and system availability is lacking in \gls{bess} control.
    \item \gls{minlp} approaches for realistic and nonlinear \gls{bess} control remain underexplored due to computational complexity, and existing methods rarely incorporate industrial frameworks that emulate assets using high-fidelity simulations.
\end{enumerate}

\begin{table*}[!t]
 \scriptsize
\centering
\caption{Literature review of recent peer-reviewed papers optimizing battery power setpoints aimed at improving \gls{bess} performance and lifespan.}
\label{tab:synthesis_bullets}
\begin{tabular}{
  >{\raggedright\arraybackslash}p{1.5cm}  
  >{\raggedright\arraybackslash}p{4.8cm}  
  >{\raggedright\arraybackslash}p{4.8cm}  
  >{\raggedright\arraybackslash}p{4.8cm}  
}
\toprule
\textbf{Reference} & \textbf{Methodology} & \textbf{Contribution} & \textbf{Limitation} \\
\midrule

Yu et al. (2022)~\cite{yu2022hierarchical} & Proposes a hierarchical distributed coordinated control strategy for \gls{bess}  participating in Automatic Generation Control (AGC). The upper layer uses fuzzy control to distribute AGC commands between \gls{bess} and traditional generators based on \gls{bess} \gls{soc}. The lower layer uses a consistency algorithm for power allocation among distributed BESS units. & 
\begin{itemize}[nosep, leftmargin=*]
    \item Develops a two-level control system that utilizes the fast response of \gls{bess} for frequency regulation while considering generator characteristics and BESS \gls{soc}.
    \item Avoids excessive power consumption and \gls{soc} depletion in the BESS.
\end{itemize} & 
\begin{itemize}[nosep, leftmargin=*]
    \item Employs fixed proportion droop control for \gls{soc} and lacks in adaptivity.
    \item Focused solely on frequency regulation.
    \item No consideration of thermal and derating effects during operation.
\end{itemize} \\
\addlinespace

Hasan et al. (2022)~\cite{hasan2022multi} & Proposes a multi-objective co-design optimization framework using NSGA-II. Optimizes sizing and technology selection of grid-connected hybrid \gls{bess} with different chemistries and architectures. Objectives: total cost of ownership (TCO), efficiency, and lifetime. & 
\begin{itemize}[nosep, leftmargin=*]
    \item Framework enables detailed techno-economic design of HBESS.
    \item Demonstrates a 29.6\% TCO reduction.
    \item Supports mixed chemistries and architectures.
\end{itemize} & 
\begin{itemize}[nosep, leftmargin=*]
    \item Cell-level lumped mass thermal models.
    \item No consideration of thermal impact during operation and thermal derating at the system level.
    \item Nonlinearities were ignored in modeling battery efficiency.
\end{itemize} \\
\addlinespace

Li et al. (2022)~\cite{li2022optimal} & Formulates a multi-objective BESS siting/sizing optimization using an Improved Equilibrium Optimizer (IEO). Objectives: LCC, power loss, peak-shaving cost, tie-line fluctuation, and voltage deviation. & 
\begin{itemize}[nosep, leftmargin=*]
    \item Captures a wide range of planning criteria.
    \item IEO improves solution quality over standard EO.
\end{itemize} & 
\begin{itemize}[nosep, leftmargin=*]
    \item Focuses on planning, not real-time control.
    \item Thermal and inverter losses are not modeled.
    \item No consideration of spatial temperature distribution and thermal derating.
\end{itemize} \\
\addlinespace

Song et al. (2023)~\cite{Song.2023} & The paper presents a multi-objective optimization framework to schedule BESS operation in a virtual power plant (VPP), targeting cost reduction and voltage regulation. Applied to a real low-voltage network in Alice Springs, the tool integrates with DIgSILENT PowerFactory for realistic, multi-time scale simulations. & 
\begin{itemize}[nosep, leftmargin=*]
    \item Balances customer cost and grid impact using Pareto-optimal charge/discharge schedules over 24 hours.
    \item Tested in DIgSILENT PowerFactory on real low-voltage networks across multiple time scales.
    \item Performs well under varying battery sizes and renewable generation levels.
\end{itemize} & 
\begin{itemize}[nosep, leftmargin=*]
    \item The tool targets voltage regulation and cost efficiency but does not account for \gls{bess} thermal dynamics or battery aging.
    \item The method uses offline workflow  and is unsuitable for real-time BESS control.
    \item Optimization is performed at customer level rather than for multi-string BESS used in grid or industrial settings.
\end{itemize} \\

\addlinespace

Pant et al. (2023)~\cite{pant_soc} & Proposes a real-time power-sharing algorithm for BESS racks based on both SOC and temperature. The control scheme dynamically allocates power set-points by accounting for SOC disparity and thermal deviation, without requiring complex SOH estimation. Designed to enhance battery longevity, especially in second-life BESS deployments. & 
\begin{itemize}[nosep, leftmargin=*]
    \item Introduces a generic temperature-SOC control algorithm suitable for real-time grid applications.
    \item Reduces reliance on SOH estimation, simplifying control for heterogeneous second-life batteries.
    \item Demonstrates thermal management efficacy, contributing to battery life extension.
\end{itemize} & 
\begin{itemize}[nosep, leftmargin=*]
    \item Focuses primarily on thermal and \gls{soc}-based degradation, without explicit multi-objective economic modeling.
    \item No explanation for optimization weight selection.
    \item No consideration of battery and inverter power derating
\end{itemize} \\

\addlinespace

Jin et al. (2023)~\cite{Jin.2024} & The study models a multi-carrier energy system and applies multi-objective optimization to generate a Pareto frontier of scheduling strategies. Instead of relying on subjective weights, battery \gls{soh}, computed via physical modeling, is used as an automated \gls{kpi} to select the optimal solution. & 
\begin{itemize}[nosep, leftmargin=*]
    \item Multi-objective optimization incorporating battery \gls{soh} as a decision-making indicator.
    \item Generates a complete Pareto frontier considering environmental and economic objectives.
\end{itemize} & 
\begin{itemize}[nosep, leftmargin=*]
    \item It lacks a real-time control implementation, limiting its use for dynamic BESS operation.
    \item Simple lumped mass temperature estimation.
    \item No consideration of battery power derating.
\end{itemize} \\

\addlinespace

Tanjavooru et al. (2025)~\cite{Tanjavooru.2025} &  Proposes an optimal control strategy for SOC balancing and a framework to analyze spatial temperature distribution in multi-pack BESS. & 
\begin{itemize}[nosep, leftmargin=*]
    \item Integrates thermal simulation and SOH estimation with power split control strategies.
    \item Compares \gls{mpc} and Rule-Based Control for power sharing in BESS.
    \item Demonstrates \gls{mpc}’s advantage in maintaining uniform and cooler temperature profiles and reducing battery aging.
\end{itemize} & 
\begin{itemize}[nosep, leftmargin=*]
    \item No feedback of accurate states from simulation.
    \item \gls{lp} based optimizer for only single objective, \gls{soc} balancing.
    \item Framework currently limited to estimate spatial temperature and \gls{soh}, without active awareness of full system-level dynamics. 
\end{itemize} \\
\addlinespace

Cornejo et al. (2025)~\cite{cornejo2025evaluating} & Compares an \gls{lp} vs. a high-fidelity \gls{nlp} model using ECM for intraday energy arbitrage. Simulates across battery aging scenarios. & 
\begin{itemize}[nosep, leftmargin=*]
    \item Shows \gls{nlp} model yields better revenue and efficiency.
    \item Captures power limitations in aged batteries.
\end{itemize} & 
\begin{itemize}[nosep, leftmargin=*]
    \item Assumes perfect \gls{soc}/\gls{soh} knowledge.
    \item Focuses on revenue, not broader trade-offs.
    \item No consideration of spatial temperature distribution and thermal derating.
\end{itemize} \\

\addlinespace

Zhao et al. (2025)~\cite{zhao2025hierarchical} & Introduces a Hierarchical and Self-Evolving Digital Twin (HSE-DT). A Transformer-CNN deep learning model estimates battery state (SOC, SOH) using historical and real-time data. Transfer learning enables model adaptation as the battery ages. & 
\begin{itemize}[nosep, leftmargin=*]
    \item Builds a robust DT framework for battery awareness.
    \item Achieves SOC RMSE less than 0.9\%.
    \item Integrates multiple models in a self-evolving hierarchy.
\end{itemize} & 
\begin{itemize}[nosep, leftmargin=*]
    \item Needs evaluation under real-time and extreme conditions.
    \item Focused on state estimation, not optimization/control.
\end{itemize} \\

\bottomrule
\end{tabular}
\end{table*}

\subsection{Paper contributions}
\begin{enumerate}
\item Although \gls{minlp} approaches exist in literature, this work is, to the authors’ knowledge, the first to apply an \gls{minlp}-based power split control for a multi-string \gls{bess} that incorporates temperature and \gls{soc} dependent efficiency and derating loss models.
\item The developed power split control is integrated with a high-fidelity electro-thermal simulation of the \gls{bess}, enabling real-time acquisition of accurate battery states to establish \gls{mpc} control.
\item A Pareto-based multi-objective analysis highlights the inherent trade-off between battery and inverter efficiency: minimizing battery losses reduces derating but increases inverter losses, and vice versa. A best-compromise solution is identified that balances these competing objectives while maintaining system availability and efficiency. 

\end{enumerate}

\section{Methodology}
This control methodology follows the following sequence: the optimization framework is introduced, outlining the interactions between the components required to establish the \gls{mpc} control $\rightarrow$ the hierarchical and blended objective definitions are explained $\rightarrow$ the mathematical formulations of nonlinear constraints within the optimization are described $\rightarrow$ the \glspl{kpi} used to evaluate the optimal control performance are defined.

\begin{figure}[!htbp]
    \centering
    \includegraphics[width=1.09\linewidth]{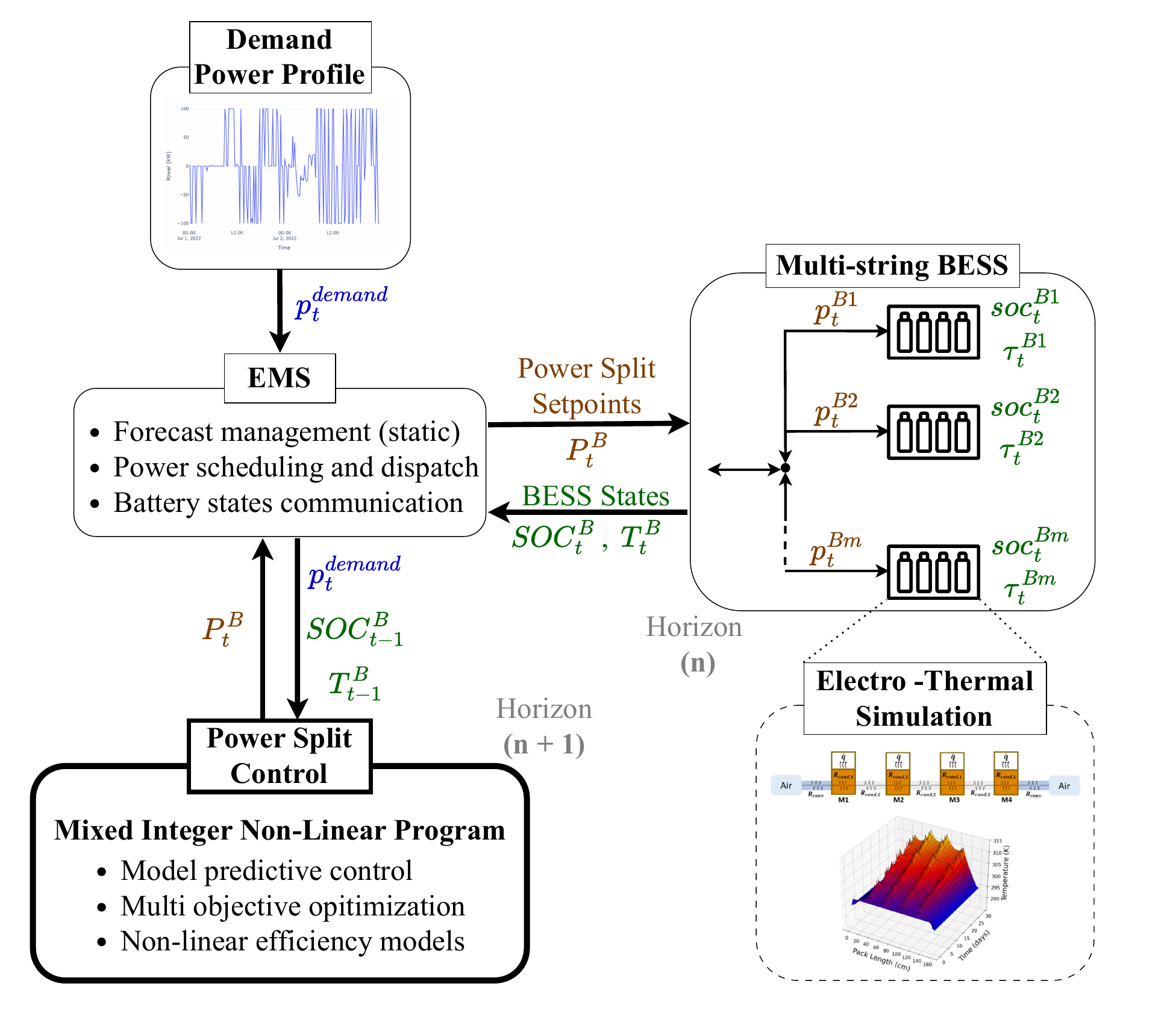}
    \caption{Framework showcasing the interaction of Power Split Control with \gls{ems} and \gls{bess} with power demand (blue), optimal power setpoints (brown), and battery states (green).}
    \label{fig:framework}
\end{figure}


\subsection{Optimization Framework}

The proposed control strategy is formulated using \gls{minlp} that incorporates electro-thermal dynamics and derating effects of \gls{bess} into the optimization layer. This optimization model is designed to maximize overall system availability while minimizing inverter losses, battery losses, and temperature overshoot in the battery, which leads to derating losses. The framework shown in Fig. \ref{fig:framework} represents a realistic industrial control and monitoring architecture for a multi-string \gls{bess} operating under dynamic conditions driven by a load profile. Each \gls{bess} string comprises an inverter and the battery string. In this work, static wholesale market data, which serves as the input to the \gls{bess}, is read by the \gls{ems}, which is modeled to create a rolling horizon of data for the \gls{mpc} based power split control. The \gls{ems} processes this profile to determine optimal energy dispatch strategies. These strategies are passed to the Power Split Control module, which is the focus of this work. This \gls{minlp} optimization-based control allocates power commands across individual battery strings. The \gls{bess} states obtained from the simulation in the current horizon (n) are used to re-initialize the states in the power split control for the next optimization horizon (n+1). Each string in the multi-string \gls{bess} is monitored for its states \gls{soc} and temperature. To support thermal-aware control, a high-fidelity Electro-Thermal Simulation is integrated, comprising an \gls{ecm} and a \gls{fdm} based 1D spatial thermal simulation that evaluates internal temperature distributions based on battery power and ambient air cooling~\cite{Tanjavooru.2025}. This closed-loop architecture enables updation of the battery states for the next \gls{mpc} optimization horizon.

\subsection{Objective Function}

In this work, we addressed the challenge of managing trade-offs among multiple conflicting objectives by utilizing a combination of hierarchical (lexicographic) and blended objective handling. Each objective is assigned a priority \( P_i \) reflecting its structural importance, while weights \( W_i \) enable exploration of Pareto-optimal trade-offs among lower-priority objectives. The blended formulation uses a weighted sum of all objectives, which captures the assigned priorities and tunable weights. Hierarchical control ensures that lower-priority objectives do not degrade the optimal values of higher-priority ones, thereby maintaining critical system constraints while systematically exploring trade-offs in other objectives~\cite{GurobiOptimization.2024}.
\begin{subequations}
\begin{align}
\text{Priority levels} \quad & P_1 > P_2 > \dots > P_n  \label{eq:priority_levels} \\[2pt]
\min_{x} \quad & F_{P_k}(x) \quad \text{lexicographically w.r.t.\ } P_k  \label{eq:lex_min} \\[4pt]
\text{where} \quad & F_{P_k}(x) = \sum_{\substack{i \\ P_i = P_k}} W_i\,\cdot f_i(x) \label{eq:blended} \\[6pt]
\text{subject to} \quad &F_{P_j}(x) = F_{P_j}^* + \varepsilon_1, \quad \forall\, P_j > P_k  \label{eq:preserve_higher_priority}
\end{align}
\end{subequations}

After determining the optimal value \( F_{P_1}^* \) for the highest-priority objective \( F_{P_1}(x) \), the solver minimizes the next objective \( F_{P_2}(x) \), subject to the constraint \( F_{P_1}(x) = F_{P_1}^* + \varepsilon_1\), where a small tolerance (\(\varepsilon_1\)) is introduced to relax the constraint and allows numerical rounding. This process iterates over all objectives, preserving the optimality of higher-priority goals.

\begin{table}[!htbp]

\centering
\caption{Priority order and weights for \gls{moo}.}
\begin{tabular}{l c c}
\hline
\textbf{Objective} & \textbf{Priority (P)} & \textbf{Weight (W)} \\ \hline
Availability Loss  & $P_1$ & $W_1$ 
\\
Derating Loss      & $P_2$ & $W_2$ 
\\
Inverter Loss      & $P_3$ & $W_3$ \\ 
Battery Loss       & $P_4$ & $W_4$\\ \hline

\end{tabular}
\label{tab:weights_priorities}
\end{table}

In the course of this work, the \gls{moo} objective functions are defined as described in Table.~\ref{tab:weights_priorities}. The objective function accounts for losses arising from limited available capacity for charge/discharge, inverter switching, battery heat generation, and power derating required for safe operation. Minimizing this objective function enables efficient \gls{bess} dispatch while adhering to battery \gls{soc} and temperature constraints.

\subsection{Power Balance and SOC Constraints}
For all $t \in T$ time and $m \in M$ strings:

\begin{equation}
0 \leq p^{B[m]}_t \leq p^{N}
\label{eq:power_limit}
\end{equation}
The power delivered by each battery string, $ p^{B[m]}_t $ is limited by its inverter's nominal power, $p^{N}$, as defined in \eqref{eq:power_limit}.

\begin{equation}
\sum_{m \in M}p^{B[m]}_t = p^{demand}_t
\label{eq:power_total}
\end{equation}
The summation of powers of all the parallel battery strings is constrained to meet the forecasted power demand, $p^{demand}_t$, as per \eqref{eq:power_total}.

\begin{equation}
\begin{split}
p^{B[m],ch}_t + p^{inv[m]}_t + p^{heat[m]}_t + {} \\
p^{avail,high[m]}_t + p^{derate,loss[m]}_t = p^{B[m]}_t + {} \\
s^{B[m],+}_t - s^{B[m],-}_t  \\
\end{split}
\label{eq:power_balance_1}
\end{equation}

During charging scenarios, the power balance accounts for multiple operational losses, including battery loss (heat), inverter loss, temperature-based derating loss, and availability loss during fully charged conditions as shown in \eqref{eq:power_balance_1}. These losses are collectively considered as components of the battery string power flow. Positive and negative slack variables are introduced to relax this equality constraint.  
\begin{equation}
\begin{split}
p^{B[m],dch}_t - p^{inv[m]}_t - p^{heat[m]}_t + {} \\
p^{avail,low[m]}_t + p^{derate,loss[m]}_t = p^{B[m]}_t + {} \\
s^{B[m],+}_t - s^{B[m],-}_t
\end{split}
\label{eq:power_balance_2}
\end{equation}

During discharge scenarios, the losses from heat and inverter are met by the discharge power as shown in Eq.~\ref{eq:power_balance_2}. This implies a higher discharge power to meet the required battery string power. And the availability loss during discharge arise from the lack of usable charge in the battery.
\begin{equation}
0 \leq soc^{B[m]}_t \leq 1
\label{eq:soc_limits}
\end{equation}
The operating limits of the \gls{soc} of each battery string are constrained as per \eqref{eq:soc_limits}.
\begin{equation}
soc^{B[m]}_{t} = soc^{B[m]}_{t-1} + \frac{\Delta t}{E^{N}} \cdot 
(I^{B[m],ch}_{t} - I^{B[m],dch}_{t})
\label{eq:coloumb}
\end{equation}
The \gls{soc} estimation inside the optimization layer follows a coulomb counting methodology considering the charge and discharge battery currents as defined in \eqref{eq:coloumb} \cite{Mohammadi.2022}.

\subsection{Availability Loss Estimation}
\begin{equation}
b^{high[m]}_{t} = 1 \Rightarrow soc^{B[m]}_{t} \geq 1 - \epsilon
\label{eq:b_high_1}
\end{equation}
\begin{equation}
b^{high[m]}_{t} = 0 \Rightarrow soc^{B[m]}_{t} \leq 1 - \epsilon
\label{eq:b_high_0}
\end{equation}
\begin{equation}
b^{low[m]}_{t} = 1 \Rightarrow soc^{B[m]}_{t} \leq \epsilon
\label{eq:b_low_1}
\end{equation}
\begin{equation}
b^{low[m]}_{t} = 0 \Rightarrow soc^{B[m]}_{t} \geq \epsilon
\label{eq:b_low_0}
\end{equation}

Two binary variables are introduced to model the conditional constraints at the battery's upper and lower limit operation based on its \gls{soc} as shown in \eqref{eq:b_high_1}--\eqref{eq:b_low_0}. When each of these binary variables return 1, as in \eqref{eq:p_avail_high}--\eqref{eq:p_avail_low}, the corresponding losses associated with the availability of the concerned string will be non-zero . 
\begin{equation}
p^{avail,high[m]}_t \leq M^{avail} \cdot b^{\text{high}[m]}_t + \epsilon
\label{eq:p_avail_high}
\end{equation}
\begin{equation}
p^{avail,low[m]}_t \leq M^{avail} \cdot b^{\text{low}[m]}_t + \epsilon
\label{eq:p_avail_low}
\end{equation}

To ensure that availability losses are activated in the power balance constraints only at either the upper or lower \gls{soc} limits, the two binary indicator variables are constrained to be mutually exclusive as shown in \eqref{eq:binaries}.
\begin{equation}
b^{high[m]}_{t} + b^{low[m]}_{t} \leq 1
\label{eq:binaries}
\end{equation}

\subsection{Battery Thermal Model}
Within the optimizer, a lumped mass thermal model is used to estimate the temperature evolution of the battery string as defined in \eqref{eq:lumped_thermal}. After each horizon of optimization, the temperature estimations of the string are corrected using a high fidelity spatial thermal simulation as shown in Fig.~\ref{fig:framework} \cite{Tanjavooru.2025}. This way, each string's accurate mean temperature is updated at every optimization step to reflect realistic thermal dynamics, enabling predictive thermal control.
\begin{equation}
\tau^{B[m]}_{t} = \tau^{B[m]}_{t-1} + \Delta t \cdot
\left(k1 \cdot p^{heat[m]}_{t}
- k2 \cdot (\tau^{B[m]}_{t-1} - \tau_{\text{air}}) \right)
\label{eq:lumped_thermal}
\end{equation}

Since the optimizer in this work also aims to minimize battery losses, the heat generation due to internal resistance was modeled in its nonlinear form, grounded in the underlying physical principles for greater accuracy as described below. \\

\subsubsection{Internal Resistance Model}

To capture realistic system behavior, this study integrates temperature-dependent and \gls{soc}-dependent internal resistance modeling into the optimization layer. 
\begin{equation}
b^{soc[m]}_{t} = 1 \Rightarrow soc^{B[m]}_{t} \leq 0.1 + \epsilon^{soc}
\label{eq:soc_low_1}
\end{equation}
\begin{equation}
b^{soc[m]}_{t} = 0 \Rightarrow soc^{B[m]}_{t} \geq 0.1 - \epsilon^{soc}
\label{eq:soc_low_2}
\end{equation}

While the variation of internal resistance with temperature is gradual, it increases abruptly with \gls{soc} below 0.1~\cite{Doh.2019}. In order to account for this behavior, a binary variable is introduced to switch between the behaviors at \gls{soc} equals 0.1. To relax this switching constraint a small tolerance ($\epsilon^{soc}$) is introduced as shown in \eqref{eq:soc_low_1}--\eqref{eq:soc_low_2}.

\begin{equation}
b^{soc[m]}_{t} = 1 \Rightarrow R^{1[m]}_t \;=\; f\!\bigl(soc^{B[m]}_{t}\bigr) \cdot \frac{f\!\bigl(\tau^{B[m]}_{t}\bigr)}{2.5} \quad
\label{eq:R1}
\end{equation}
\begin{equation}
b^{soc[m]}_{t} = 0 \Rightarrow R^{2[m]}_t \;=\; f\!\bigl(\tau^{B[m]}_{t}\bigr) \quad
\label{eq:R2}
\end{equation}

When the binary variable is active (i.e., equals 1), the internal resistance is modeled as a function of both \gls{soc} and temperature with the temperature influence normalized relative to its highest value as in \eqref{eq:R1}. Conversely, when the binary variable is inactive (i.e., equals 0), the resistance is modeled only as a function of temperature as the value of \gls{soc} has minimal influence as defined in \eqref{eq:R2}~\cite{Doh.2019}. 

\begin{equation} 
f\!\bigl(soc^{B[m]}_{t}\bigr): LUT\;=\;
\begin{cases}
40\,\mathrm{m}\Omega, \hspace{1cm} \quad soc^{B[m]}_{t} = 0,\\[4pt]
u_i\,soc^{B[m]}_{t} + v_i, \\
\quad \qquad soc_i <soc^{B[m]}_{t} \le soc_{i+1}, \\ 
\hspace{2cm} \quad  (i = 1,\dots,n\!-\!1) \\[4pt]
2.5\,\mathrm{m}\Omega, \hspace{1cm} soc^{B[m]}_{t} = 0.1
\end{cases}
\label{eq:pwl_resistance_SOC}
\end{equation}

\begin{equation} 
f\!\bigl(\tau^{B[m]}_{t}\bigr): LUT\;=\;
\begin{cases}
2.5\,\mathrm{m}\Omega, & \tau^{B[m]}_{t} \le 25\,^\circ\mathrm{C},\\[4pt]
c_i\,\tau^{B[m]}_{t} + d_i, 
& \tau_i <\tau^{B[m]}_{t} \le \tau_{i+1}, \; \\
& \quad (i = 1,\dots,n\!-\!1) \\[4pt]
0.75\,\mathrm{m}\Omega, & \tau^{B[m]}_{t} = 80\,^\circ\mathrm{C}
\end{cases}
\label{eq:pwl_resistance_T}
\end{equation}

The nonlinear dependencies on \gls{soc} and temperature are modeled as piecewise linear approximations using  \glspl{lut} over the relevant operating ranges as shown in \eqref{eq:pwl_resistance_SOC} and \eqref{eq:pwl_resistance_T} respectively. The combined effect of these dependencies is modeled as the total internal resistance ($ R^{B[m]}_t $) used for calculating heat generation in this work as per \eqref{eq:total_R}. 
\begin{equation}
R^{[m]}_t = 
R^{1[m]}_t\cdot b^{soc[m]}_{t}
+ R^{2[m]}_t \cdot \big(1 - b^{soc[m]}_{t}\big)
\label{eq:total_R}
\end{equation}

\subsubsection{Equivalent Circuit Model}
Rint equivalent circuit model is modeled into the optimization layer to derive currents based on charge and discharge powers \cite{He.2011}. The current is computed as the root of a quadratic equation derived from the relationship between \gls{ocv}, internal resistance and the requested charge/ discharge powers as shown in \eqref{eq:ecm}. In order to derive \gls{ocv}, the nonlinear \gls{soc}-\gls{ocv} relationship is approximated as an \gls{lut} differently for both charge and discharge scenarios \cite{Baccouche.2017, Kucevic.2020b}. 

\begin{equation}
 I^{[m]}_t =  f\!\bigl(R^{[m]}_{t}, OCV^{[m]}_{t}, P^{B[m],ch/dch}_{t}\bigr)
 \label{eq:ecm}
\end{equation}

Heat generation is subsequently calculated as a function of the derived current and the total internal resistance as per \eqref{eq:p_heat}. 
\begin{equation}
 p^{heat[m]}_t =  {I^{[m]}_t}^2. R^{[m]}_t
 \label{eq:p_heat}
\end{equation}
\subsection{Temperature-Based Battery Derating}

Temperature-based battery derating mimics real-world battery safety mechanisms by reducing the allowable power as battery temperature increases. \gls{li-nmc} batteries have an optimal operational range of 15--45\,\si{\celsius}, beyond which performance and capacity degradation emerge \cite{Feinauer.2024}.

\begin{equation} 
k^{derate}_t: LUT\;=\;
\begin{cases}
1 & \tau^{B[m]}_{t} \le 45\,^\circ\mathrm{C},\\[4pt]
x_i\,\tau^{B[m]}_{t} + y_i, 
& T_i <\tau^{B[m]}_{t} \le T_{i+1}, \; \\
& \quad (i = 1,\dots,n\!-\!1) \\[4pt]
0 & \tau^{B[m]}_{t} \ge 60\,^\circ\mathrm{C}
\end{cases}
\label{eq:pwl_derating}
\end{equation}

From the simulation model, the battery string's thermal sensitivity, considering air cooling, was found to be approximately 0.012\,\si{\celsius} per kilowatt of power input over a 15-minute control interval. This indicates a sluggish thermal response, which demands a gradual control strategy to derate the input power.

\begin{figure}[!htbp]
    \centering
    \includegraphics[width=\linewidth]{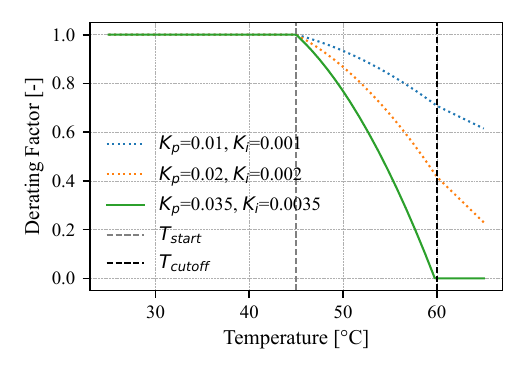}
    \caption{Temperature-based derating factor vs. battery temperature.}
    \label{fig:derating}
\end{figure}

Instead of relying on a fixed linear derating curve \cite{Barreras.2018}, this work implements a \gls{pi} controller that dynamically adjusts the derating factor ($ k^{derate}_t $) in response to temperature deviations from a set threshold. The proportional term reacts to immediate deviation, while the integral term accounts for heat accumulation over time, enabling smoother yet responsive control. The controller gains \(K_p\) and \(K_i\), as shown in Fig.~\ref{fig:derating}, were tuned using a trial-and-error approach to ensure that the battery temperature remained within the safe operational limits under various load conditions. The simulation trials showed that the highlighted gains of 0.035 and 0.0035 for \(K_p\) and \(K_i\) respectively were sufficient to prevent overheating, considering system’s slow thermal dynamics along with ambient air cooling and intermittent inactivity.  In the optimizer, this nonlinear derating factor ($ k^{derate}_t $) as shown in Fig.~\ref{fig:derating} is approximated as a piecewise linear \gls{lut} as defined in  \eqref{eq:pwl_derating} and the derating loss is computed as per \eqref{eq:p_derate}.

\begin{equation}
p^{derate,loss[m]}_t =k^{derate}_t \cdot p^{B[m]}_t
\label{eq:p_derate}
\end{equation}

\subsection{Inverter Loss Model}

Inverter loss is also modeled using Big M method set of constraints. As shown in \eqref{eq:bigM1}--\eqref{eq:bigM3} a large constant \(M^{\mathrm{inv}}\) and a small tolerance \(\epsilon^{\mathrm{inv}}\) link the battery power \(p^{B[m]}_t\) with the binary variable \(b^{\mathrm{inv}[m]}_t\). 
\begin{equation}
p^{B[m]}_t \leq M^{inv} \cdot b^{inv[m]}_{t} + \epsilon^{inv}
\label{eq:bigM1}
\end{equation}
\begin{equation}
b^{inv[m]}_{t} = 0 \Rightarrow  p^{inv[m]}_t \leq \epsilon
\label{eq:bigM2}
\end{equation}
\begin{equation}
b^{inv[m]}_{t} = 1 \Rightarrow  p^{inv[m]}_t: LUT
\label{eq:bigM3}
\end{equation}

When the inverter is “off” (\(b=0\)), this forces the battery power to (approximately) zero; when “on” (\(b=1\)), the inverter output power \(p^{\mathrm{inv}[m]}_t\) is modeled to follow the \gls{lut} as shown in \eqref{eq:pwl_inverter}.
\begin{equation}
 p^{inv[m]}_t: LUT =
\begin{cases}
0, & p^{B[m],ch(/dch)}_t = 0, \\[4pt]
a_i\,p^{B[m]}_t + b_i, 
& P_i < p^{B[m]}_t \le P_{i+1}, \\
c_i\,p^{B[m],dch}_t + d_i, 
& P_i < p^{B[m],dch}_t \le P_{i+1}, \\ 
& \quad (i = 1,\dots,n\!-\!1) \\[4pt]
0.0193 \cdot p^{N}, & p^{B[m],ch}_t = p^{N}\\
0.0177 \cdot p^{N}, & p^{B[m],dch}_t = p^{N}
\end{cases}
\label{eq:pwl_inverter}
\end{equation}

The piecewise data for modeling the inverter losses \gls{lut} is based on experiments for a commercial inverter system in both charge and discharge scenarios.

\subsection{Key Performance Indicators}
In this work, the solution of the \gls{moo} optimal control is characterized by the following system \glspl{kpi} \eqref{eq:availability}--\eqref{eq:system_efficiency}, expressed in terms of the optimization model variables:

\begin{equation}
\text{Availability (\%)} 
= 1 
- \frac{\displaystyle\sum_{t}\sum_{m}p^{\mathrm{avail,high}[m]}_{t}+ p^{\mathrm{avail,low}[m]}_{t}}
       {\displaystyle\sum_{t}\sum_{m}p^{demand}_{t}}
       \label{eq:availability}
\end{equation}

\begin{equation}
\text{Derating Efficiency (\%)} 
= 1 
- \frac{\displaystyle\sum_{t}\sum_{m}p^{\mathrm{derate,loss}[m]}_{t}}
       {\displaystyle\sum_{t}\sum_{m}p^{B[m]}_{t}}
       \label{eq:derating_eff}
\end{equation}

\begin{equation}
\text{Inverter Efficiency (\%)} 
= 1 
- \frac{\displaystyle\sum_{t}\sum_{m}p^{\mathrm{inv}[m]}_{t}}
       {\displaystyle\sum_{t}\sum_{m}p^{B[m]}_{t}}
       \label{eq:inv_eff}
\end{equation}

\begin{equation}
\text{Battery Efficiency (\%)} 
= 1 
- \frac{\displaystyle\sum_{t}\sum_{m}p^{\mathrm{heat}[m]}_{t}}
       {\displaystyle\sum_{t}\sum_{m}p^{B[m]}_{t}}
       \label{eq:battery_eff}
\end{equation}

\begin{equation}
\text{System Efficiency (\%)} 
= 1 
- \frac{\displaystyle\sum_{t}\sum_{m} p^{\mathrm{inv}[m]}_{t} + p^{\mathrm{heat}[m]}_{t}}
       {\displaystyle\sum_{t}\sum_{m}p^{B[m]}_{t}}
       \label{eq:system_efficiency}
\end{equation}

System efficiency measures the proportion of delivered battery power that is not lost to inverter conversion losses and battery heat generation losses. Derating losses are left out in this calculation because they are intentionally imposed curtailments for thermal or operational safety, not inherent energy losses. They vary according to how conservative the proportional and integral gains are modeled for the control. Inverter and battery losses represent true irrecoverable inefficiencies within the \gls{bess}. By isolating these physical losses, system efficiency remains a pure metric of conversion performance.  

In summary, the complexity of the proposed methodology arises from the \gls{minlp} technique, which integrates modeling of availability loss, inverter loss, internal resistance, battery loss and the equivalent circuit model. In the next section, the developed methodology is applied to two case studies to analyze \gls{moo} performance. Case 1 highlights the implementation of hierarchical control with availability loss minimization set as the primary objective. And Case 2 deals with a Pareto analysis for establishing trade-offs between competing objectives. The aim is to implement the proposed methodology to evaluate the controller’s ability to balance a realistic industrial \gls{bess} availability, derating, and efficiency.

\section{Case 1: Hierarchical \gls{moo} of Availability} 

In Case 1, the \gls{moo} seeks to maximize \gls{bess} availability while minimizing the losses listed in Table \ref{tab:case1}. To ensure fair comparison, all objectives are min–max normalized to the range \([0,1]\), preventing any single metric from dominating due to unit or magnitude differences.  

\begin{itemize}
  \item \textbf{Availability Loss} (\(P=2,\;W=1\)) is treated as the top‐priority objective and is minimized first, guaranteeing maximum system uptime.  
  \item \textbf{Derating Loss}, \textbf{Inverter Loss}, and \textbf{Battery Loss} (\(P=1,\;W=1\)) share a lower priority. Once availability has been optimized, these three efficiency‐related objectives are collectively minimized, subject to a small rounding tolerance that preserves the availability optimum.
\end{itemize}

By structuring the objectives lexicographically, the solver enforces that critical availability targets are met before exploring trade‐offs among secondary efficiency goals, resulting in a balanced and reliable control strategy.

\begin{table}[!htbp]
\centering
\caption{Case 1: Objective function definition.}
\begin{tabular}{c c c c c }
\hline
\textbf{S.No.} & \textbf{Objective} & \textbf{$P_i$} & \textbf{$W_i$} & Remarks \\ \hline
1 & Availability Loss  & 2 & 1& High Priority \\ 
2 & Derating Loss      & 1 & 1& Normal \\  
3 & Inverter Loss      & 1 & 1& Normal \\ 
4 & Battery Loss       & 1 & 1& Normal  \\ \hline
\end{tabular}
\label{tab:case1}
\end{table}

\subsection{Results}
In this work, the developed \gls{minlp} based control is applied on a two-string homogeneous \gls{bess} to clearly showcase the evolution trends of \gls{soc} and temperature with higher clarity. As discussed in the previous section, the \glspl{kpi} used to assess the performance of this \gls{moo} control using the objective function described in the combination of Table.~\ref{tab:case1} are listed below:

\begin{figure}[h!]
    \centering
    \includegraphics[width=\linewidth]{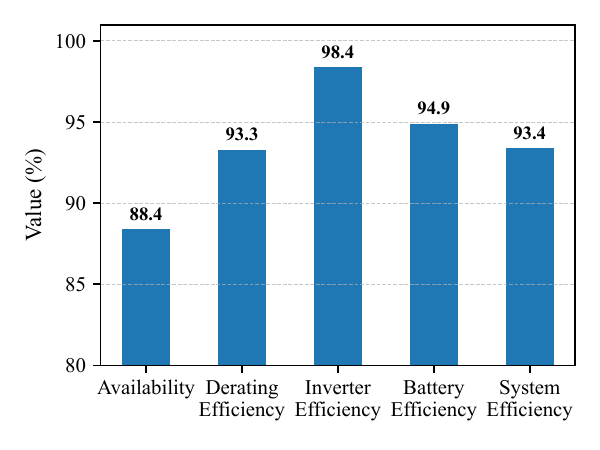}
    \caption{Performance values of \glspl{kpi} in Case 1: Availability as the primary objective with uniform weights assigned to all objectives.}
    \label{fig:histogram}
\end{figure}

In this scenario, both the battery strings have uniform initial conditions with \glspl{soc} 0.5 and temperature equal to ambient temperature of  \SI{25}{\degreeCelsius}. It is to be noted that the optimization is focused on minimizing losses and therefore has an influence in creating non-uniformity of \gls{soc} and temperature during operation.

\begin{figure}[h!]
    \centering
    \includegraphics[width=\linewidth]{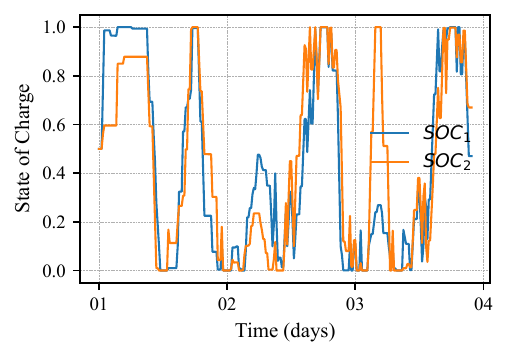}
    \caption{\gls{soc} evolution of two string system for Case 1.}
    \label{fig:SOC_MOO}
\end{figure}

The \gls{soc} trajectories indicate that, despite identical initial \glspl{soc}, the controller might have preferentially dispatched individual strings to maximize inverter efficiency which led to \glspl{soc} imbalance as shown in Figure~\ref{fig:SOC_MOO}. By avoiding partial‐load operation, this strategy reduces associated partial load losses and improves overall system performance~\cite{Schimpe.2018b}. 

\begin{figure}[h!]
    \centering
    \includegraphics[width=\linewidth]{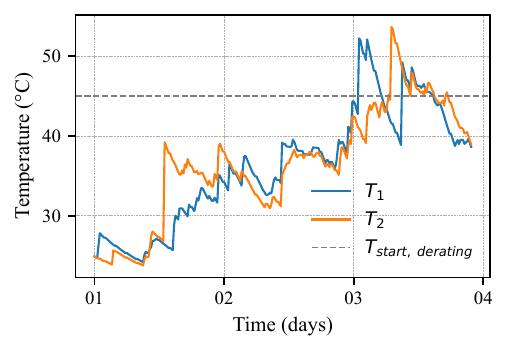}
    \caption{Temperature evolution of two string system for Case 1.}
    \label{fig:T_MOO}
\end{figure}

Although individual string dispatch improves inverter efficiency, prioritizing battery efficiency with an equal weight results in the simultaneous operation of all strings, promoting uniform temperature trajectories and minimizing battery losses caused by \(I^{2}R\) heat generation ~\cite{Tanjavooru.2025}. This behavior is evident in the temperature profiles during operation, as illustrated in Figure~\ref{fig:T_MOO}. These results suggest that prioritizing inverter efficiency may cause \gls{soc} imbalance, while prioritizing battery efficiency helps achieve temperature balancing. However, the impact of these competing objectives can only be fully understood through a Pareto analysis. Another key consideration is the derating loss, which is triggered at \SI{45}{\degreeCelsius}. To mitigate excessive temperature rise, the power output from each string is curtailed, thereby limiting internal heat generation and slowing down temperature growth.

\section{Case 2: Pareto Analysis}
A closer inspection of the optimization results from Case 1 suggests an inherent conflict between inverter and battery efficiency. Because improvements in one metric often come at the expense of the other, a Pareto analysis is needed. By tracing the Pareto frontier between these two objectives, we can identify knee points where a modest sacrifice in one efficiency yields a large gain in the other, guiding the selection of a balanced operating strategy for the BESS. This sections aims to conduct a Pareto analysis by varying the weights of the inverter and battery losses as shown in the Table.~\ref{tab:pareto_analysis}. The remaining weights and the priorities are maintained similar to Case 1.

\begin{table}[!htbp]
\centering
\caption{Priorities and Weight Sweep Used in Pareto Analysis.}

\vspace{1em}
\begin{tabular}{c c c c l}
\hline
\textbf{Objective} & \textbf{$P_i$} & \textbf{$W_i$} & \textbf{Range} & \textbf{Remarks} \\
\hline
Availability Loss & 2 & 1 & Fixed    & High Priority \\
Derating Loss     & 1 & 1 & Fixed    & Normal \\
Inverter Loss     & 1 & [0, 1] & Variable & Pareto Sweep \\
Battery Loss      & 1 & [0, 1] & Variable & Pareto Sweep \\
\hline
\end{tabular}

\vspace{2em}
\textbf{Weight Sweep Scenarios for $W_3$ and $W_4$}

\vspace{0.5em}
\begin{tabular}{l c c c}
\hline
 Weights & \textbf{S1} & \textbf{S2} & \textbf{S3} \\
\hline
$W_3$ & 1   & 0.5 & 0 \\
$W_4$ & 0   & 0.5 & 1 \\
\hline
\end{tabular}

\label{tab:pareto_analysis}
\end{table}

In this case, three optimization scenarios (S1--S3) were formulated to perform a Pareto analysis, as detailed in Table~\ref{tab:pareto_analysis}. Scenario S1 minimizes only the inverter loss, disregarding the battery loss objective. In contrast, Scenario S3 focuses solely on minimizing battery loss, excluding inverter loss from the objective. Scenario S2 assigns partial weights to both inverter and battery losses, representing a trade-off configuration. Notably, derating loss is assigned a fixed weight of 1 across all scenarios, ensuring it remains a consistently prioritized objective throughout the analysis.

\subsection{Results}

\begin{figure}[h!]
    \centering
    \includegraphics[width=\linewidth]{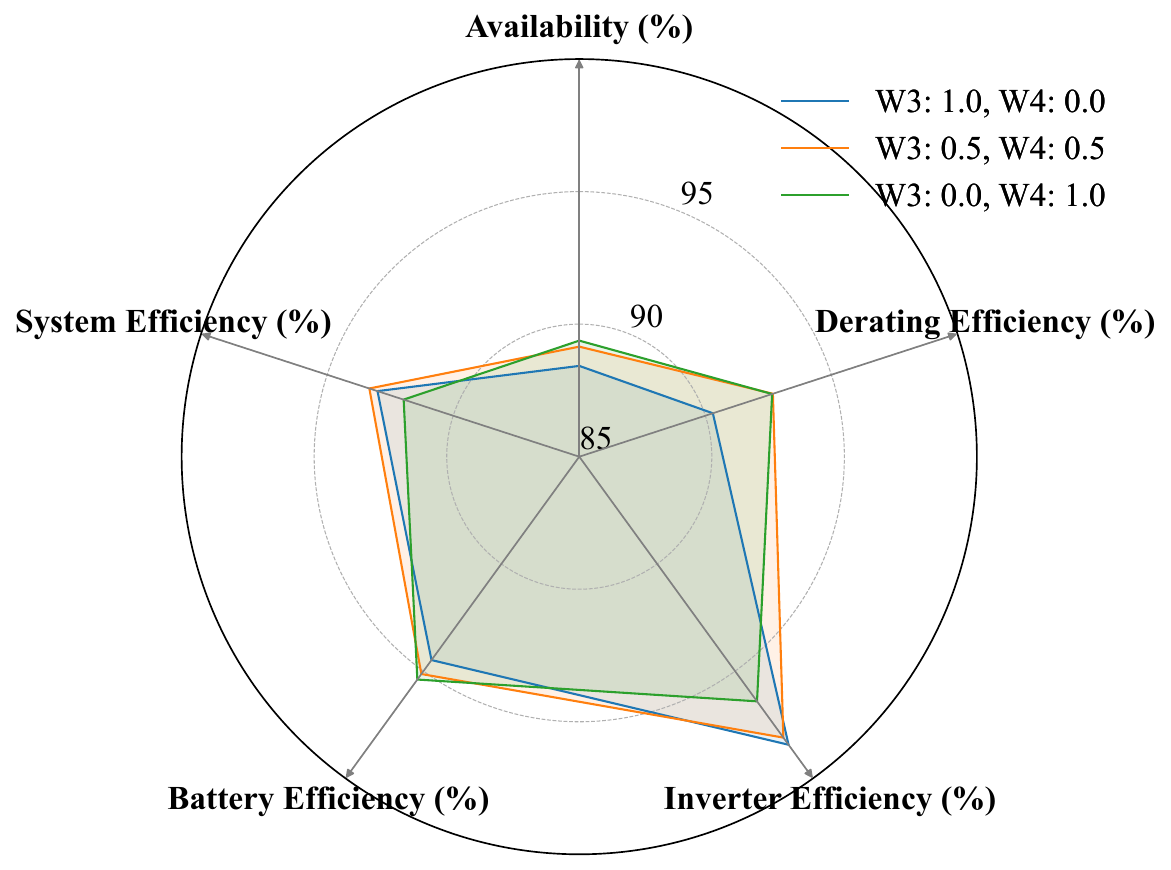}
    \caption{Pareto Analysis for inverter (W3) and battery (W4) losses in Case 2.}
    \label{fig:pareto}
\end{figure}

Fig.~\ref{fig:pareto} illustrates the radar plot comparing the \glspl{kpi} of the three optimization scenarios (S1--S3) conducted as part of the Pareto analysis.

\begin{itemize}
  \item \textbf{Scenario S1} (\textcolor{blue}{W3: 1.0, W4: 0.0}) emphasizes minimizing inverter losses exclusively. This results in the highest inverter efficiency, as power is concentrated in fewer strings operating near their nominal power. However, the battery efficiency and derating performance degrade due to elevated heat generation from concentrated currents.

  \item \textbf{Scenario S2} (\textcolor{orange}{W3: 0.5, W4: 0.5}) represents a balanced trade-off between inverter and battery efficiency. While it does not achieve the extreme values observed in S1 or S3 for either inverter or battery efficiency, it delivers a well-rounded performance across all \glspl{kpi}. Notably, it achieves the highest system efficiency among all scenarios and maintains strong performance in derating efficiency, making it a robust compromise between conflicting objectives. 

  \item \textbf{Scenario S3} (\textcolor{OliveGreen}{W3: 0.0, W4: 1.0}) prioritizes the minimization of battery losses. By distributing power evenly across all strings, it maximizes battery efficiency through the reduction of \( I^2R \) heat. Additionally, by avoiding thermal hot-spots in individual strings, this scenario limits peak temperature events, thereby improving derating efficiency as well. However, this load balancing causes inverters to operate away from their optimal power point, leading to reduced inverter efficiency. With respect to system availability, this uniform dispatch results in more homogeneous thermal and SOC profiles, which enhances overall system availability \cite{Tanjavooru.2025}. 

\end{itemize}

\section{Discussions}

The findings from Case~1 and Case~2 highlight the value of multi-objective optimization in achieving reliable and efficient BESS operation. In Case~1, a fixed-priority framework was used where availability loss had the highest priority, ensuring that system availability was never compromised. This configuration provided balanced performance across inverter, battery, and derating efficiencies, although it limited the exploration of trade-offs among these competing objectives. Case~2, on the other hand, employed a Pareto analysis by varying the weights between the inverter and battery loss objectives. The results clearly demonstrated a conflict: Scenario~S1 maximized inverter efficiency but incurred higher battery losses and reduced derating efficiency due to concentrated current flow; Scenario~S3 minimized battery losses by distributing current across strings, achieving higher battery and derating efficiencies but lower inverter efficiency. Scenario~S2 emerged as a balanced solution, offering the highest system efficiency while maintaining strong performance across all key performance indicators. 
\begin{figure}[h!]
    \centering
    \includegraphics[width=\linewidth]{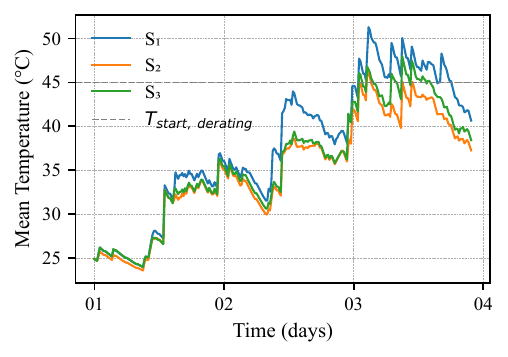}
    \caption{Plot of mean BESS temperature of S1--S3 scenarios in Case 2.}
    \label{fig:T_allCases}
\end{figure}

Another key insight from this analysis is the impact of temperature-based power derating. Figure~\ref{fig:T_allCases} shows the mean temperature trajectories of the \gls{bess} across scenarios (S1--S3). Scenario~S1 exhibits the highest temperature rise of \SI{51}{\degreeCelsius}, resulting in greater derating losses. However, the extent of derating is influenced by the derating temperature \( T_{\text{start}} = \SI{45}{\degreeCelsius} \) and the selected \gls{pi} controller gains (\( K_p \) and \( K_i \)). These parameters, which govern the control responsiveness, can be tuned depending on the operational priorities of the \gls{bess} such as minimizing degradation or increasing availability. By adjusting these variables, it is possible to mitigate derating losses and thereby enhance system availability further.

\section{Conclusion}

This work presented a \gls{minlp} optimization model embedded in an \gls{mpc} framework to achieve state-aware, multi-objective optimization of an air-cooled \gls{bess} considering temperature-based derating.  The controller receives real-time state updates from a high-fidelity electro-thermal simulation, ensuring that the optimization reflects the true internal conditions of each battery string.  Two case studies validated the approach.  \textbf{Case 1} introduced a lexicographic plus blended objective formulation that simultaneously maximized availability and minimized inverter, battery and derating losses without compromising critical up-time targets. \textbf{Case 2} applied a Pareto analysis to vary the weighting between inverter and battery efficiency objectives, revealing clear trade-offs and identifying a balanced operating point that achieved the highest overall system efficiency. The balanced operation achieved efficiency improvements of 1\% in the battery, 1.5\% in the inverter, and 2\% in thermal derating, while sustaining high system availability. Moreover, a peak temperature reduction of \SI{5}{\degreeCelsius} in the \gls{bess} indicates lower thermal stress without compromising performance and availability. Collectively, these results confirm that the proposed \gls{minlp}–\gls{mpc} framework effectively manages competing objectives thereby ensuring compliance with critical constraints for safety, and responds dynamically to real-time \gls{bess} conditions for more reliable and efficient operation. By explicitly accounting for both electrical and thermal intricacies of the system, this framework provides a solid foundation for advanced, multi-domain control architectures in future grid-integrated storage systems with evolving \gls{bess} applications.

\section{Future scope of work}

Future work will focus on enhancing the flexibility, scalability, and real-world applicability of the proposed control framework through several key developments:

\begin{itemize}
\item \textbf{New objectives:} The \gls{moo} can be extended to include long-term metrics such as battery aging and economic cost, enabling cost-optimal dispatch over the battery’s lifetime.

\item \textbf{System inhomogeneities:} Incorporating initial \(\Delta SOC\) and \(\Delta T\) across strings can evaluate the controller’s ability to reduce uneven degradation thereby optimizing for efficiency.

\item \textbf{Granular Pareto analysis:} A finer Pareto analysis will explore trade-offs among objectives, which can be used to generate a dynamic \gls{lut} for real-time decision making in \gls{bess} control.

\item \textbf{Cooling strategy evaluation:} The framework will be extended to assess advanced cooling methods such as liquid and hybrid cooling and their impact on thermal dynamics and derating.


\item \textbf{Computational benchmarking:} A comparative study for linear vs nonlinear control models will be conducted to evaluate whether faster feedback updates compensate for higher fidelity modeling to better suit the real-time performance.
\end{itemize}

\section{Acknowledgment}
This manuscript is based on research conducted within the project KI-M-Bat: Artificial Intelligence-Based Modular Battery Systems for Industrial and Grid Applications, funded by the Bayerische Forschungsstiftung (BFS). The authors would like to thank FENECON GmbH for providing battery system data used for parameterization and simulation validation in this work.

\section{Data Availability}
The complete framework and code will be made openly available for reproducibility and future extensions. The battery parameterization data are available subject to request.

\bibliographystyle{unsrt}
\bibliography{ref}

\end{document}